\documentclass[useAMS,usenatbib]{emulateapj}
\usepackage{latexsym,graphicx,natbib}

\def\simless{{\th \rlap{\raise 0.5ex\hbox{$\scriptstyle  {<}$}}
    {\lower 0.3ex\hbox{$\scriptstyle  {\sim}$}} \th }}  
\def\simgreat{{\th \rlap{\raise 0.5ex\hbox{$\scriptstyle  {>}$}}
    {\lower 0.3ex\hbox{$\scriptstyle  {\sim}$}} \th }}  
\def\greateq{{\th \rlap{\raise 0.5ex\hbox{$\scriptstyle  {>}$}}
    {\lower 0.3ex\hbox{$\scriptstyle  {-}$}} \th }}  
\def\lesseq{{\th \rlap{\raise 0.5ex\hbox{$\scriptstyle  {<}$}}
    {\lower 0.3ex\hbox{$\scriptstyle  {-}$}} \th }}  
\def\th{\thinspace}
\def\ts{{\raise 0.3ex\hbox{$\scriptstyle {\th \sim \th }$}}}

\usepackage{epsfig}
\newcommand{\etal}{\mbox{et al.}}
\newcommand{\xtejfull}{\mbox{XTE J1807$-$294}}

\newcommand{\xtej}{\mbox{XTE J1807}}

\newcommand{\nudot}{\dot{\nu}}

\begin{document}

\shortauthors{Patruno \etal}
\shorttitle{Coherent timing  of \xtej\ }

\title{Accretion torques and  motion of the 
hot spot on the accreting millisecond pulsar XTE J1807-294 }
\author{Alessandro Patruno \altaffilmark{1}, Jacob M. Hartman\altaffilmark{2}, R. Wijnands \altaffilmark{1},
Deepto Chakrabarty\altaffilmark{3}, Michiel van der Klis \altaffilmark{1} }
\altaffiltext{1}{Astronomical Institute ``Anton Pannekoek,'' University of
Amsterdam, Kruislaan 403, 1098 SJ Amsterdam, The Netherlands, E-mail: a.patruno@uva.nl}
\altaffiltext{2}{Space Science Division, Naval Research Lavoratory, Washington, DC 20375, USA}
\altaffiltext{3}{Department of Physics and Kavli Institute for Astrophysics
and Space Research, Massachusetts Institute of Technology, Cambridge, MA
02139, USA}

\email{a.patruno@uva.nl}

\begin{abstract}

We present a coherent timing analysis of the 2003 outburst of the
accreting millisecond pulsar \xtejfull\ . We find an upper limit for
the spin frequency derivative of $|\dot{\nu}|< 5\times
10^{-14}\rm\,Hz/s$. The sinusoidal fractional amplitudes of the
pulsations are the highest observed among the accreting millisecond
pulsars and can reach values of up to 27$\%$ (2.5-30 keV). The pulse arrival time
residuals of the fundamental follow a linear
anti-correlation with the fractional amplitudes that suggests hot spot
motion over the surface of the neutron star both in longitude and
latitude. An anti-correlation between residuals and X-ray flux
suggests an influence of accretion rate on pulse
phase, and casts doubts on the use of standard timing techniques to
measure spin frequencies and torques on the neutron star.

\end{abstract}

\section{Introduction}

An open problem in the field of accreting millisecond pulsar (AMXP) is
how to devise a reliable method to measure spin and orbital
parameters.  Since the discovery of the first AMXP \citep{Wijnands98}
considerable improvements have been made, leading to the measurement
of accurate orbital and spin parameters for 9 of the 10 known AMXPs
(see~\citealt{Wijnands04},~\citealt{Poutanen06},~\citealt{DiSalvo07} for a
review and~\citealt{Patruno09} for the last source
discovered). Current methods (see e.g.~\citealt{Taylor92}) are based on
folding procedures to reconstruct the pulse profiles of the accreting
neutron star and on direct measurement of the pulse phase variations
due to orbital Doppler shift and spin changes (for example due to
torques). The pulse phases are fitted using $\chi^{2}$ minimization
techniques. However, a substantial complication sometimes arises due
to the presence of a strong unmodeled noise component in the pulse
phases that, when ignored, might affect the reliability of the
method. Two possible strategies have been used in the literature to
try and overcome this: (i) harmonic data selection (\citealt{Burderi06},~\citealt{Riggio08},~\citealt{c08},~\citealt{Papitto07})
and (ii) use of a minimum variance estimator (\citealt{BD85},~\citealt{har08}).  In the first case the pulse profiles are
decomposed into their harmonic components: generally one sinusoid at
the fundamental frequency (or first harmonic, $\nu$) and one at the
second harmonic ($2\nu$) and are analyzed separately, measuring two
independent sets of orbital and spin parameters.  The harmonic with
the weakest noise content is selected for the measurement of the spin
and orbital parameters and the noisier one is discarded (see e.g.,~\citealt{Burderi06}).  Although this use of the most ``stable''
harmonic reduces the $\chi^{2}$, this selection throws away part of
the information and in that sense is not optimal. The hypothesis
behind the selection of the most stable harmonic is that, for unknown
reasons, that harmonic better tracks the spin of the neutron star.
\citet{Burderi06} speculated that the second harmonic might be
more stable because it arises from accretion onto both the polar caps
and hence is insensitive to the flux ratio between poles.

In the second method, both harmonics are used and weighted to minimize
the effect of phase noise (\citealt{BD85},~\citealt{har08}). However,
also in this second situation in practice data selection is
performed. If the phases of both harmonics change differently, the
possibility of defining pulse arrival times breaks down and the data
where this happens have to be excluded from the analysis
\citep{har08}.\\ Because both methods employ different data
selections, different results are obtained when analyzing the same
source. For example in the case of SAX J1808.4--3658, the pulse
frequency derivative $\nudot$, measured from only the second harmonic
was $4.4\times 10^{-13} \rm\, Hz\, s^{-1}$ for the first 14 days of
the 2002 outburst and $-7.6\times 10^{-14} \rm\, Hz\,s^{-1}$ for the
rest of the outburst \citep{Burderi06}. In \citet{har08} we considered
the same source and gave an upper limit of $|\nudot|<2.5\times
10^{-14} \rm\, Hz\,s^{-1}$ for all the four outbursts for which high
resolution timing data was available. The reason for this discrepancy
is that while \citet{Burderi06} used only the information carried by
the second harmonic and rejected the results of the fundamental
frequency, we used both harmonics but excluded the initial data where
the phase variations where stronger and discrepant between harmonics
\citep{har08}.  So these differences arise as a consequence of
different data selections.
In this paper we try to better characterize the timing noise such as
observed in AMXPs focusing on a source where the noise is strong: XTE
J1807-294 (J1807 from now on) which has been in outburst for $\approx
120$ days in 2003 \citep{Markwardt03}.

\section{Data reduction and reconstruction of the pulse profiles}
\label{technique}

We reduced all the pointed observations from the {\it{RXTE}} satellite
taken with the Proportional Counter Array (PCA, \citealt{Jahoda06})
that cover the 2003 outburst of J1807. The PCA instrument provides an
array of five proportional counter units (PCUs) with a collecting area
of 1200 $\rm\,cm^{2}$ per unit operating in the 2-60 keV range and a
field of view with a FWHM of $\sim 1^{\circ}$. 

We constructed the X-ray lightcurve using the counts in PCA
Absolute channels 5-67 ($\approx 2.5-29$ keV).  

We constructed our pulse profiles by folding 512 s long chunks of
lightcurve in profiles of $N=32$ bins, with the ephemeris of
\citet{Riggio08}. In this folding process we used the TEMPO pulsar
timing program to generate a series of polynomial expansions of the
ephemeris that predict the barycentered phase of each photon
detected. The total number of photons detected in a single profile bin
is $x_{j}\pm\sqrt{x_{j}}$, with the error calculated from counting
statistics and $j=1,...,N$.  Since the pulse profile shape changes
throughout the outburst, it is not possible to base the analysis on a
stable template profile. Therefore we decided to analyze the pulse
profile harmonic components separately.

To calculate the pulse fractional amplitudes and phases we decomposed
each profile as:
\begin{equation}
x_{j}=b_{0}+\displaystyle\sum_{k}b_{k}\rm\,cos\left\{\it 2\pi\left[\frac{k(j-0.5)}{N} - \phi_{k}\right]\right\}
\end{equation}

by using standard $\chi^{2}$ minimization techniques.  The term
$b_{k}$ is the amplitude of the sinusoid representing the $k-$th
harmonic, and $b_{0}$ is the unpulsed flux component.  We choose the
first peak of each sinusoid in the profile as the fiducial point for
each harmonic.  Defining the $k$-th harmonic frequency to be
$k\cdot \nu$, the unique pulse phases $\phi_{k}$ of each harmonic
range from 0 to 1.  The i--th pulse time of arrival (TOA) of the k--th
harmonic is then defined as: $t_{k,i}= \frac{\phi_{k}}{k\cdot
\nu}+\Delta t_{i}$.  Here $\Delta t_{i}$ is the time of the middle of
the i-th folded chunk.  With these definitions, a positive time shift
is equivalent to a lagging pulse TOA, while a negative shift
corresponds to a preceding pulse TOA.  This is the convention that
will be used later to define pulse phase residuals.

The fractional sinusoidal amplitude of the i-th pulse profile and the k-th
harmonic is calculated as:
\begin{equation}
R_{i,k}= \frac{N\times b_{k}}{N_{ph,i}-B_{i}}
\end{equation}
where $\rm N_{\it ph,i}$ and $\rm B_{\it i}$ are the total number of
photons and the background counts (calculated with the FTOOL
\emph{pcabackest}) in the i-th pulse profile. 
The error on the
fractional amplitude $R_{i,k}$ is calculated propagating the errors on
$b_{k}$ and $N_{\it ph,i}$.  The error
on $B_{i}$ is negligible with respect to the other errors and will not be
considered further.

We define a pulse profile harmonic to be significant if the ratio
between the amplitude $b_{k}$ and its statistical error
$\sigma_{b_{k}}$ is larger than 3.3 when using a folding time of 512 s. 
The choice of 3.3 guarantees that
the number of false detections expected when considering the
global number of pulse profiles ($\approx 850$), is less than one.
The length of the folding time was then changed to 300 and $3000$ s to
probe different timescales (see \S 3), and the significance threshold
rescaled to 3.5 and 3$\sigma$ respectively, according to the new
number of pulse profiles.

After obtaining our set of TOAs for all the significant harmonics we
chose to describe the phase $\phi$ of the k-th harmonic (we omit the k
index from now on) at the barycentric reference frame, as a
combination of six terms:
\begin{eqnarray}\label{eq:phi}
\phi\left(t\right) = \phi_{L}\left(t\right) +  \phi_{Q}\left(t\right) + \phi_{O}\left(t\right) +
\phi_{M}\left(t\right)\nonumber\\
 +  \phi_{A}\left(t\right) + \phi_{N}\left(t\right) 
\end{eqnarray}
where $\phi_{L}\left(t\right)$ is a linear function of the time
($\phi_{L}\left(t\right)=\phi_{0}+\nu\,t$, with $\phi_{0}$ an initial
reference phase), $\phi_{Q}\left(t\right)$ is a parabolic function of
time ($\phi_{Q}\left(t\right)=\frac{1}{2}\nudot t^{2}$), and
$\phi_{O}\left(t\right)$ is the keplerian orbital modulation
component.  The term $\phi_{M}\left(t\right)$ is the measurement error
component, and is given by a set of independent values and is normally
distributed with an amplitude that can be predicted by propagating the
Poisson uncertainties due to counting statistics.  The term
$\phi_{A}\left(t\right)$ is the astrometric uncertainty position
error, and the last term, $\phi_{N}\left(t\right)$, is the so-called
timing noise component that defines all the phase variations that
remain.  The timing noise includes, but is not limited to, any phase
residual that can be described as red noise and possible extra white
noise in addition to that described by the measurement error component
$\phi_{M}\left(t\right)$.

One of the key points when dealing with timing noise is how to
distinguish a true spin frequency change of the neutron star from an
effect that mimics it. In general, $\phi_{Q}$ and $\phi_{N}$ can both
be due to torques, both \emph{not} be due to torques, or one can,
while the other is not.  In the first case the torque is not constant
and has a fluctuating component. In the second case there is a process
different from a torque affecting the pulse phases. In the third case,
if $\phi_{Q}$ is due to a torque, it is constant, while if $\phi_{N}$
is due to a torque then the torque is not constant.

In the presence of timing noise ($\phi_{N}$) the formal parameter
errors estimated using standard $\chi^{2}$ minimization techniques are
not realistic estimates of the true uncertainties, as the hypothesis
behind the $\chi^{2}$ minimization technique is that the source of
noise is white and its amplitude can be predicted from counting
statistics.  In the presence of an additional source of noise, such as
the timing noise, the apparently significant measurement of a
parameter can simply reflect the non realistic estimation of the
parameter errors. To solve this, we adopted the technique we already
employed in \citet{har08}, who used Monte
Carlo (MC) simulations of the timing residuals to account for the
effect of timing noise on the parameter errors.  The technique uses
the power density spectrum of the best-fit timing residuals of a
$\nudot$ model, as output by TEMPO. Then thousands of fake power
density spectra are produced, with Fourier amplitudes identical to the
original spectrum and with random uncorrelated Fourier phases. The
Fourier frequencies are then transformed back to the time domain into
fake residuals, and thousands of $\nu$ and $\nudot$ values are
measured to create a Gaussian distribution of spin frequencies and
spin frequency derivatives. The standard deviations of these
distributions are the statistical uncertainties on the spin frequency
and derivative.  For a detailed explanation of the method we refer to
\citet{har08}.

\section{Results}
\subsection{Measurement of the spin frequency and its derivative in the presence of timing noise}\label{par:spin}

 We fitted the phases of each harmonic with a circular keplerian model
($\phi_{O}$) plus a linear term ($\phi_{L}$) and a quadratic term
($\phi_{Q}$).  All the residual phase variation we observe after
removing these three terms is treated as noise ($\phi_{M}$ and
$\phi_{N}$).  The $\nu$ and $\nudot$ measured for each of the two
harmonics is given in Tables 1 and 2, respectively.  The errors on the
pulse frequency and its derivative are calculated performing $10^{4}$
MC simulations as described in \S~\ref{technique}.  At long periods
(days), red noise dominates the power spectrum, while at short periods
(hours), the uncorrelated Poisson noise dominates.  The red noise
power spectrum is not very steep, and has a power law dependence
$P(\nu)\propto \nu^{\alpha}$ with $\alpha\approx -0.5$.

The source position we used comes from {\it Chandra} observations
whose $68\%$ confidence level error circle is $0''.4$ in radius.  The
astrometric uncertainty introduced in this way on the frequency and
frequency derivative is $3\times 10^{-8}\rm\, Hz$ and $0.7\times
10^{-14}\rm\, Hz\,s^{-1}$, respectively (calculated with eqs. A1 and A2
from \citet{har08}, which added in quadrature to the MC
statistical errors gives the final errors reported in Tables 1 and 2.
The final pulse frequency derivative significances for the fundamental
and the second harmonic are $\approx 2.7\sigma$ and $\approx
1.5\sigma$, respectively.

We note that the significance of the frequency derivative for the
fundamental increases above the 3 sigma level when the statistical
errors are calculated with standard $\chi^{2}$ minimization
techniques, consistently with \citet{Riggio08}.  These errors
calculated with $\Delta\chi^{2}=1.0$ are $2\times 10^{-16}\rm\,
Hz\,s^{-1}$ and $1.6\times 10^{-15}\rm\, Hz\,s^{-1}$ for the
fundamental and second harmonic respectively.  So, a significant
$\nudot$ is present which is, however, consistent with being
part of the (red) timing noise.

\begin{figure}[t]
  \begin{center}
    \rotatebox{-90}{\includegraphics[width=0.7\columnwidth]{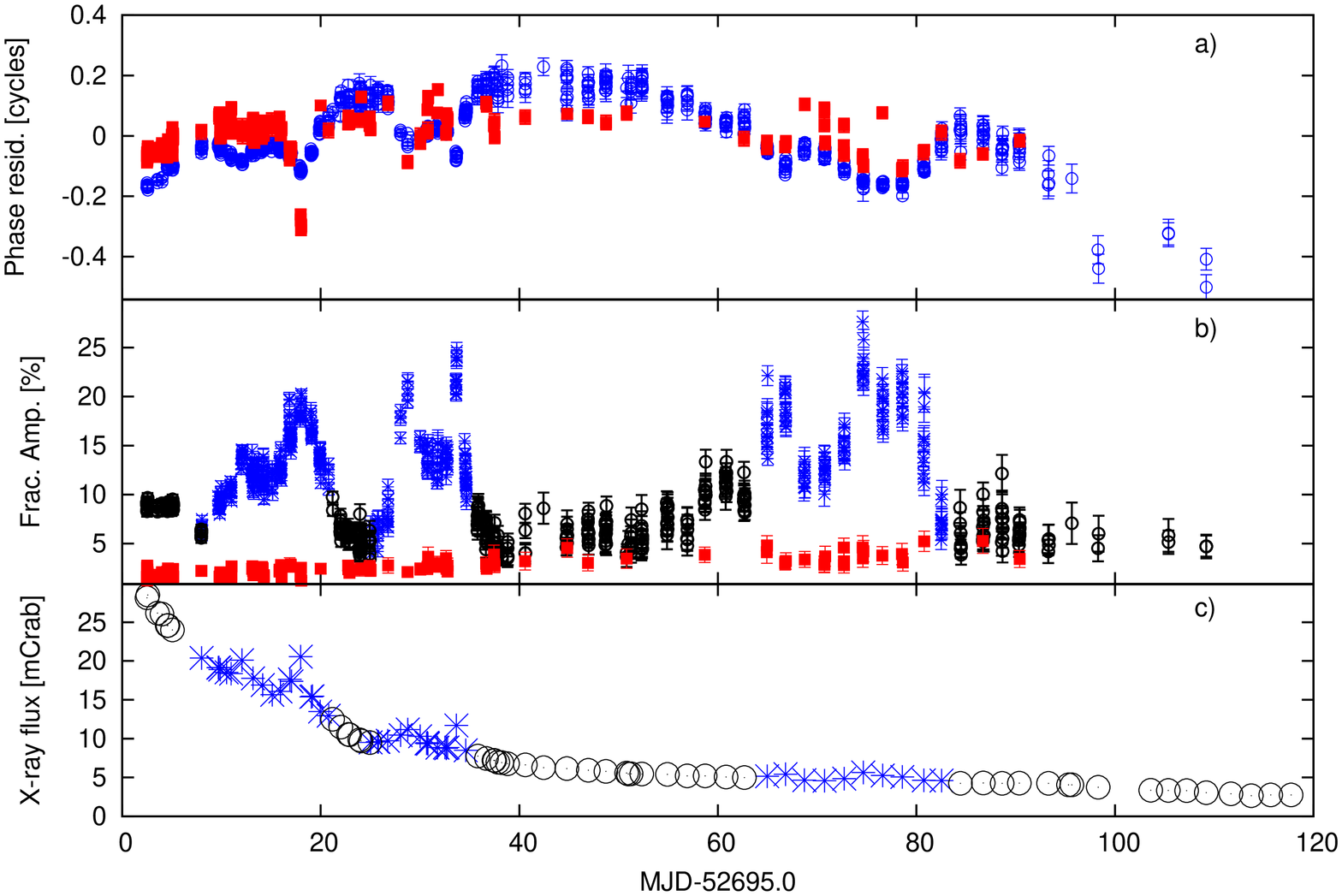}}
  \end{center}
  \caption{ \textbf{a):} Timing residuals for a constant spin
frequency and a circular keplerian orbit. The fundamental (blue
circles) and the second harmonic (red squares) phases were measured
using an integration time of $512$ s per pulse profile.\\ \textbf{b):}
Sinusoidal fractional amplitude of the fundamental (blue asterisks:
flaring; black circles: non-flaring) and second harmonic (red squares)
during the whole outburst. During the flares, the fundamental
sinusoidal fractional amplitude grows up to $\approx 27\%$, which is
the highest value ever observed for an AMXP.\\ \textbf{c):} \xtejfull\
lightcurve of the 2003 outburst. The count rate was normalized to the
Crab \citep{Kuulkers94} using the data nearest in time and in the
same PCA gain epoch (e.g., \citealt{Straaten03}).  The blue
circles and the black asterisks identify the 4 non-flaring and the 3
flaring states, respectively, as defined in \citet{c08}.
    \label{fig:resid}}
\end{figure}

The timing residuals obtained after removing a $\nudot=0$ model are
plotted in Figure~\ref{fig:resid} for both the harmonics (see Tables 1
\& 2 for the pulse frequencies used in the fits).  Our orbital
solution is consistent for the two harmonics and with the orbital
parameters published in \citet{Riggio08}. For the fundamental we
find:
\begin{itemize}
\item orbital period: 2404.4163(3) s
\item projected semi-major axis: 4.830(3) lt-ms
\item time of ascending node: MJD 52720.675601(3)
\end{itemize}
where the quoted errors are calculated with the $\chi^{2}$
minimization technique and correspond to $\Delta\chi^{2}=1$.  Since
the pulse phase residuals are approximately white and consistent with
the expected Poissonian uncertainty, on timescales equal to and
shorter than the orbital period, the orbital parameter
errors are a good approximation of the true uncertainties.  \\

\begin{table*}
\caption{Timing parameters for XTE J1807-294 (Fundamental)}

\scriptsize
\begin{center}
\begin{tabular}{lllll}
\hline
\hline
Parameter & Fundamental & MC error & Astrometric error & Final error\\
\hline
Spin frequency $\nu$ (Hz)  & 190.62350702 &  $2\times 10^{-8}$Hz & $3\times 10^{-8}$Hz & $4\times 10^{-8}$Hz\\
Spin frequency derivative $\dot{\nu}\rm\,(10^{-14} Hz\,s^{-1})$  & $2.7$ & $0.7$ & $0.7$ & $1.0$\\

Reference Epoch (MJD) & 52720.0 \\
\hline
\end{tabular}
\end{center}
\label{tab:timsol}
\end{table*}
\begin{table*}
\caption{Timing parameters for XTE J1807-294 (second harmonic)}

\scriptsize
\begin{center}
\begin{tabular}{lllll}
\hline
\hline
Parameter & second harmonic & MC error & Astrometric error & Final error \\
\hline
Spin frequency $\nu$ (Hz)  & 190.62350706 & $3\times 10^{-8}$Hz & $3\times 10^{-8}$Hz & $4\times 10^{-8}$Hz\\
Spin frequency derivative $\dot{\nu}\rm\,(10^{-14} Hz\,s^{-1})$  & 1.6 & $0.8$ & $0.7$ & $1.1$\\

Reference Epoch (MJD) & 52720.0 \\
\hline
\end{tabular}
\end{center}
\label{tab:timsol2}
\end{table*}

\subsection{Relation between timing residuals and X-ray flux}\label{flux-res}

In this section we analyze the relation between the pulse arrival time
residuals relative to a constant pulse frequency ($\nudot=0$) model and X-ray
flux. \citet{Riggio08} found that the residuals of the fundamental
show a strong correlation with the X-ray flux, while the second
harmonic shows only a marginal correlation.  Since large pulse phase
shifts are often observed (in both harmonics) in coincidence with the
flaring states, we investigate the possibility that at least part of
the observed timing noise is correlated with the presence of X-ray
flux variations.

\begin{figure}[t]
  \begin{center}
    \rotatebox{-90}{\includegraphics[width=0.7\columnwidth]{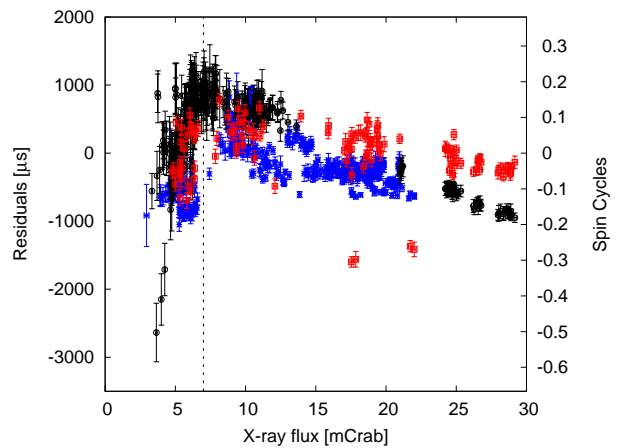}}
  \end{center}
  \caption{Phase residuals vs. X-ray flux for the fundamental (blue
asterisks for the flaring intervals and open black circles for the
non-flaring intervals) and the second harmonic (red open squares)
relative to a $\nudot=0$ model. Each pulse is a 512 s folded chunk of 
lightcurve.  The dashed line at around 7
mCrab splits the diagram in two regions: in the left one the points
follow a correlation, in the right one they follow an
anti-correlation.  The four points of the second harmonic that lay
outside the relations correspond to the large jump observed during the
second flare.
    \label{fig:fr}}
\end{figure}

In this section we show that both the harmonics are consistent with
being correlated with X-ray flux.  First we focus on the entire data
set, then we split the data in intervals choosing the same 7 chunks as
\citet{c08}; see Figure~\ref{fig:resid}c), distinguishing
\emph{non-flaring} states following the exponential flux decay of the
overall outburst, and \emph{flaring} states, comprising the six spikes
in the lightcurve. In Figure~\ref{fig:fr} we plot the residuals
vs. the count rate for both the fundamental and the second harmonic.

We applied a Spearman rank correlation test to the flux
anti-correlation for each harmonic. We accept the null hypothesis (no
correlation in the data set) if the probability $p > 1\%$.  If we do
not make any data selection, the Spearman test shows no correlation in
either harmonic. However, a clear split in the data is apparent at
around 7 mCrab: below this threshold the residuals seem to follow a
correlation with the flux, while above this threshold an
anti-correlation is visible for both harmonics. The Spearman
coefficients for the points above the threshold are $\rho=-0.8$ and
$\rho=-0.65$ for the fundamental and the second harmonic respectively
($p<1\%$). A few outliers are visible in the plot, such as for example
the four points of the second harmonic at about $\approx -0.3$
cycles. These points correspond to data taken during some of the
flaring states. If we consider only the non-flaring states, the
Spearman coefficients become $\rho=-0.9$ and $\rho=-0.76$ respectively
($p<1\%$).

The fact that we see a change from correlation to anti-correlation
around $7$ mCrabs is due to the fact that at that flux level in the
decay of the outburst the timing residuals reach the peak of the
parabolic function that dominates the residuals (at MJD$\approx
52745$, see Figure~\ref{fig:resid}).  This is a consequence of the
fitting procedure, which selects the constant reference pulse
frequency that minimizes the $\chi^{2}$ of the timing residuals.  As the
observed pulse frequency is increasing, the reference
frequency is too fast for the rising part of the residuals, and too
slow for the decreasing part.

We have seen in \S~\ref{par:spin} that the measured pulse frequency
increase is consistent with being part of a red noise process and that
true neutron star spin variations may or may not be the cause. We can
choose a higher reference pulse frequency than the one used to produce
Figure 1a, and turn the correlation-anti-correlation dichotomy in the
flux-residual diagram into only an anti-correlation, at the cost of
increasing $\chi^{2}$ by a factor $\approx 10$.  A $\nu$ higher by
$10^{-7} \rm\,Hz$ makes the split in the data disappear and increases
the degree of correlation between flux and timing residuals.

All the correlations and
anti-correlations disappear or are strongly reduced for the timing
residuals relative to the best-fit finite constant-$\nudot$ model.

\subsection{Pulse profiles}

In this section we focus on the 
shape of the pulse profiles and their relation with other 
observables, such as the phase, the timing noise and the X-ray flux.

\subsubsection{The fractional amplitude-residual diagram}

\begin{figure}[t]
  \begin{center}
    \rotatebox{-90}{\includegraphics[width=0.7\columnwidth]{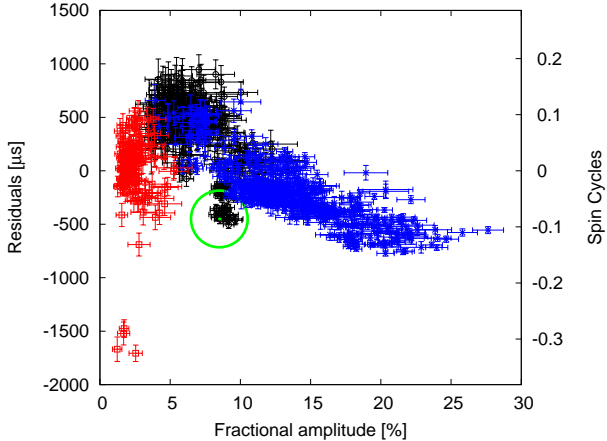}}
  \end{center}
  \caption{Timing residuals vs. fractional amplitude diagram. The blue
asterisks refer to the flaring states, while the black circles are the
non-flaring states, both referring to the fundamental frequency.  The
second harmonic is plotted as red open squares.  Each pulse was built
using 512 s of integration time. The residuals are relative to a finite
$\nudot$ model. The green circled outliers of the anti-correlation for the
fundamental, all belong to the first 2.5 days of the
observations. The second harmonic amplitude is uncorrelated to timing
residuals.
    \label{fig:far}}
\end{figure}

We have seen in the previous section that for some data selections
the X-ray flux correlates with the timing residuals relative to a
$\nudot=0$ model, but not when a finite $\dot{\nu}$ is admitted.  As
already noticed by \citet{Zhang06} and \citet{c08}, the
fractional amplitude of the pulsations shows six spikes coincident
with the six flares in the lightcurve.  Therefore, a correlation might
also exist between the fractional amplitude of the pulsations and the
arrival time residuals. Using a $\nudot=0$ model, and again using a
Spearman rank test, we found a correlation coefficient $\rho=-0.61$
($p<1\%$) for the fundamental, while no significant
correlation exists for the second harmonic.  The second harmonic is
also inconsistent with following the same correlation as the
fundamental.  Repeating the test for a $\dot{\nu}$ model we still find
no correlations for the second harmonic, but the anti-correlation
found for the fundamental becomes stronger ($\rho=-0.80$, $p<1\%$).
In Figure~\ref{fig:far} we show the fractional amplitude vs. residual
diagram (relative to a $\nudot$ model). The anti-correlation is
evident. It is interesting that the small number of points (circled
in the figure) that are outliers all belong to the first 2.5 days
of the outburst. 

We then analyzed the flaring and non-flaring states separately.  The
non-flaring state shows a weak anti-correlation with a $\nudot=0$
model ($\rho=-0.43$, $p<1\%$) which becomes slightly stronger with a
$\nudot$ model ($\rho=-0.51$, $p<1\%$) The flaring state shows an 
anti-correlation relative to a $\nudot=0$ model ($\rho=-0.58$,
$p<1\%$) that becomes much stronger for a $\nudot$ model
($\rho=-0.81$, $p<1\%$).

We found no energy dependence in this fractional
amplitude-timing residual anti-correlation (amplitude anti-correlation from
now on) when we repeated the analysis in 6 different energy bands from
2.5 to 30 keV.  The same is true for the second harmonic: no
correlation was found in any energy band.

\subsubsection{The X-ray flux and the fractional amplitude}

\begin{figure}[th!]
  \begin{center}
    \rotatebox{-90}{\includegraphics[width=0.7\columnwidth]{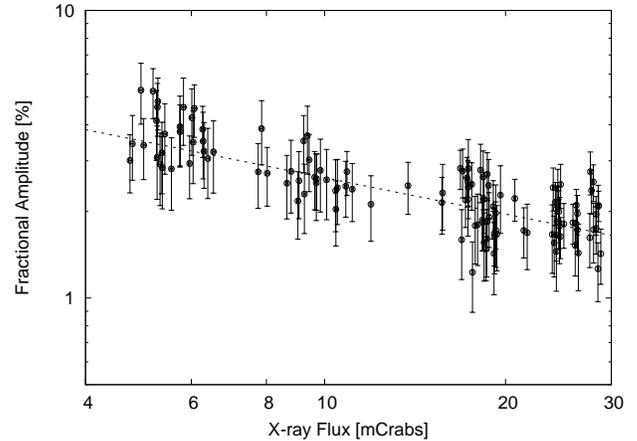}}
  \end{center}
  \caption{The fractional amplitude of the second harmonic is anti-correlated
with the flux and scales with a power law of index $\gamma=-0.41\pm 0.04$, 
close to the power law index found in a similar relation for
SAX J1808.4$-$3658.}
  \label{fig:counts}
\end{figure}

In our previous paper \citep{har08}, we found an
anti-correlation between the fractional amplitude of the second
harmonic and the X-ray flux in SAX J1808.4$-$3658.  We also noted that
the fractional amplitude of the fundamental behaved unpredictably.
Something similar applies to J1807, where no correlation is found for
the fundamental while a strong anti-correlation exists between the
observed count rate and the fractional amplitude of the second
harmonic ($\rho=-0.79$, $p<1\%$, see Figure~\ref{fig:counts}).  The
behavior of the fundamental is inconsistent with this relation.  By
analogy with \citet{har08} we fitted a simple power-law model
($R_{2}\propto f^{\gamma}_{x}$, where $f_{x}$ is the X-ray
flux) to the data, which gives a power law index $\gamma=-0.41\pm
0.04$ with a $\chi^{2}$/dof of 90.2/117.  Interestingly, the power law
index we found for SAX J1808.4$-$3658 \citep{har08} was in
agreement with this. So, a difference in behavior exists between the
fractional amplitude of the fundamental frequency and of the second
harmonic.  They respond differently to both the flux and the arrival
time residuals.

\subsubsection{Fractional amplitude}

\begin{figure}[th!]
  \begin{center}
    \rotatebox{-90}{\includegraphics[width=0.7\columnwidth]{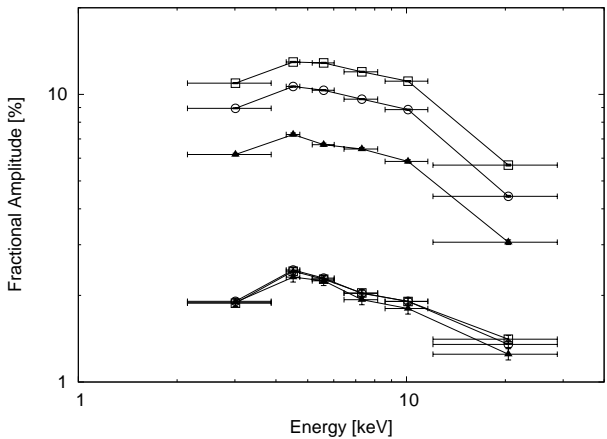}}
  \end{center}
  \caption{Energy dependence of the pulse fractional amplitudes.  The
squares and the triangles refer to the flaring and non-flaring states
respectively. The circles comprise the whole outburst. The bottom
curves, overlapped in the plot, are the fractional amplitudes of the
second harmonic which remains stable in both states.  The pulses of
the fundamental in the flaring states have a fractional amplitude
which is about 1.8 times larger than during the non-flaring states. Up
to a constant factor, the fractional amplitude has the same energy
dependence for both harmonics and for both flaring and non-flaring
states.
    \label{fig:energy}}
\end{figure}

We focus now on the energy dependence of the pulse profiles. We
consider again all the data available and the subgroups of flaring and
non-flaring states. \citet{c08} already reported on the energy
dependence of the fundamental frequency during the non flaring state.
Here we explore also the flaring state and the energy dependence of
the second harmonic.  Looking at Figure~\ref{fig:energy} two
interesting features are immediately apparent:\\ 1. the fractional
amplitude energy dependence is the same for both harmonics and 
regardless of the state of the
source (flaring, non-flaring), up to a constant factor\\ 2. the fractional amplitude of
the fundamental increases by a factor of $\approx 1.8$ during the
flaring state with respect to the non-flaring state, while it remains
approximately constant for the second harmonic.\\ Another important
property of the pulses is the time dependence of the fractional
amplitude.  In the middle panel of Figure\ref{fig:resid} we plot the
fractional amplitude of the pulsations in the $2.5-30$ keV band. As
can be seen, during the last of the six flaring states the fractional
amplitude of the fundamental increases up to $\approx 27\%$, which is
the highest ever observed for an AMXP.\footnote{In this paper we are
quoting sinusoidal fractional amplitudes, which are $\sqrt{2}$ larger than the
rms fractional amplitudes} Selecting a narrower band
between 2.5 and 10 keV the maximum fractional amplitude does not
appreciably change. During the non-flaring stage the fractional amplitude
decreases smoothly from $\approx 9\%$ down to $\approx 4\%$.
The second harmonic amplitude on the contrary increases from $\approx 2\%$ 
up to $\approx 5\%$.

\subsubsection{Harmonic content}

We decomposed each pulse profile in its harmonic components to look
for the presence of higher harmonics.  While the detection of the
second harmonic is quite common among the AXMPs, higher harmonics have
never been detected, with the exception of a possible third harmonic
in SAX J1808.4$-$3658 \citep{har08}. In J1807 we detected a
third harmonic at better than 3$\sigma$, in several different stages
of the outburst, with a maximum fractional amplitude of $\approx
1.5\%$ at MJD around 52560. To increase the S/N, we folded chunks of
data of length $3000$ s. The number of $> 3\sigma$ detections of the
third harmonic was of 11 out of 163 chunks searched.  We searched the
same chunks for a fourth harmonic, and found 5-10 significant
detections above 3$\sigma$ in the whole outburst, depending on the
binning.  When detected, the fourth harmonic has a fractional
amplitude $0.5-2.0\%$.

There were no observations where we detected all 4 harmonics at the
same time. During the second and third flares, we found a
second and fourth but not a third harmonic, during the first two
flares we found a second and third but not a fourth harmonic.  

For the third and fourth harmonics we count respectively 8 and 5
detections during the flaring states and 3 and 2 detections in the
non-flaring states.

The fractional amplitude of the third harmonic also decreases with the
flux, although the slope of the power law is much smaller
($\gamma=-0.017\pm 0.004$).  The fourth harmonic has no significant
flux dependence, but its power law slope is also consistent with the
$\gamma$ obtained for the third harmonic.

Of course this result has to be taken with caution, since we are
suffering from low number statistics with only $\approx 20$ detections
of the third and fourth harmonic altogether.

\subsection{Short-term $\dot{\nu}$ measurements}\label{sec:nudot}

Using the fundamental frequency, we measured short-term pulse
frequency derivatives using the seven sub groups of data as defined in
\S 3.2.  These measurements are useful to investigate the time
dependence of the pulse frequency derivative with time.  This test is
possible in J1807 because of its very long outburst duration (more
than 120 days, of which $\approx 106$ days with detectable
pulsations).

\begin{figure}[t]
  \begin{center}
    \rotatebox{-90}{\includegraphics[width=0.7\columnwidth]{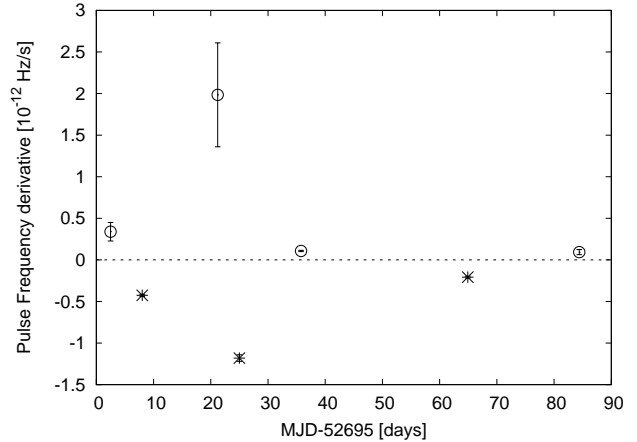}}
  \end{center}
  \caption{Frequency derivative ($\nudot$) evolution. The non flaring
states (open circles) all have positive $\nudot$, that however does
not follow a power-law decrease as expected from the accretion
theory. The flaring states (asterisks) all have a negative value,
corresponding to spin down.
\label{fig:spin-torque}}
\end{figure}

The $\nudot$ values and their uncertainties were first calculated with
standard $\chi^{2}$ minimization techniques.  All measured $\nudot$
values during the non-flaring states had a positive sign, whereas a
negative sign was measured for all three flaring states. The measured
$\nudot$ values are shown in Figure~\ref{fig:spin-torque}. There is no
clear trend, and most importantly no correlation between $\nudot$ and
the average X-ray flux in either the flaring and the non-flaring
states.  This test cannot be performed on the second harmonic, since
the smaller number of detections prevents a meaningful analysis of
data subsets for this purpose.

We then calculated the statistical uncertainties on the $\nudot$ for
each sub group of data by using the MC method as explained in
\S~\ref{par:spin}. All the $\nudot$ values were consistent with being
part of the same red noise process, consistently with what was
calculated for the long term $\nudot$ value of \S~\ref{par:spin}.

\section{Discussion}

We have analyzed the outburst of XTE J1807-294 and we have calculated
statistical errors by means of MC simulations as we previously did for
SAX J1808.4-3658 \citep{har08}. We found that with our
statistical treatment of the red noise observed in the timing
residuals of both the fundamental and the second harmonic, the
significance of the spin up is reduced below 3$\sigma$ for both the
fundamental and the second harmonic.

The fact that the spin frequency derivative is not significant does
not mean that there is not a component in the residuals that can be
fitted with a parabola. It just means that the parabola is consistent
with having the same origin as the power at other low frequencies:
both the parabola and the remaining fluctuations are consistent with
being part of the realization of the same red noise process in the
timing residuals. It is a separate issue whether or not this process
is due to true spin changes and torques on the neutron star.

Our observed parabola in the timing residuals combined with the
stochastic and astrometric uncertainty implies that any spin
frequency derivative has a magnitude smaller than $|\dot{\nu}|\simless
5\times 10^{-14}\rm\,Hz\,s^{-1}$ at the $95\%$ confidence level.

Evidence against the spin-up interpretation of the phases
comes from the lack of any correlation between the observed X-ray flux
and the measured $\nudot$ (see \S~\ref{sec:nudot}).  If standard
accretion torque theory applies, then the magnetospheric radius
($r_{m}$) should decrease as the mass accretion rate $\dot{M}$
increases, following a power-law $r_{m}\propto \dot{M}^{-\alpha}$ when
$r_{m}<r_{co}$, with $\alpha=2/7$ in the simplest case, where $r_{co}$
is the corotation radius.  This implies that also the
instantaneous $\nudot$ has a power-law dependence on the mass
accretion rate, and (when $r_{m}<r_{co}$) it is:
\begin{eqnarray}\label{torque}
\dot{\nu}&=&\frac{\dot{M}\sqrt{GMr_{m}}}{2\pi I}\simeq 1.6\times
10^{-13}\rm\,Hz\,s^{-1}\nonumber\\
&\times&\left(\frac{\dot{M}}{10^{-10}M_{\odot}\rm\,
yr^{-1}}\right)\left(\frac{\nu}{\rm\,Hz}\right)^{-1/3}\left(\frac{r_{m}}{r_{co}}\right)^{1/2}
\end{eqnarray}
 
see \citet{Bildsten97}.  Here $\dot{M}$ is the average mass
accretion rate, M the neutron star mass and $I$ the neutron star
moment of inertia.  We have observed no such a correlation between the
flux and the instantaneous pulse frequency derivative, neither in the
flaring nor in the non-flaring states. One possible explanation is
 that the X-ray flux is not a good tracer of the mass accretion rate.
If it is, standard accretion theory does not apply and the most logical
conclusion is that the observed timing residuals are not due to torques.

The possibility that the X-ray flux is not a good tracer of the mass
accretion rate is a long standing issue in the X-ray binary pulsar
field and has no simple solution. If the X-ray flux is completely
unrelated to the mass accretion rate, then no conclusions
can be drawn on the effect of the accretion on the pulse phase. 

By using eq.~(\ref{torque}) we can calculate the spin frequency
derivative expected for J1807 from standard accretion theory,
assuming a distance of 8 kpc and converting the average X-ray
luminosity into an average mass transfer rate through $L_{x}\approx
\eta c^{2}\dot{M}$.  We assume an efficiency $\eta=0.1$ for the
conversion of gravitational potential energy into radiation.  In this
way we obtain an average mass accretion rate $\dot{M}\approx 3\times
10^{-11}M_{\odot}\rm\,yr^{-1}$ (averaging over the outburst).
Assuming $r_{m}\approx r_{co}$ we have an expected $\nudot\approx
10^{-14}\rm\,Hz\,s^{-1}$, which is below our calculated upper limit of
$5\times 10^{-14}\rm\,Hz\,s^{-1}$.  However, the short term $\nudot$
values calculated in \S 3.4 exceed the theory value by 1--2 orders of
magnitude and therefore are very unlikely due to accretion
torques.

The possibility that we are not observing the effect of a torque on
the neutron star is also suggested by the fact that looking at the
shape of the lightcurve one can immediately infer the sign of the
measured pulse frequency derivative in the timing residuals.  This is
a consequence of the flux anti-correlation.  If the lightcurve is
concave, then the average $\nudot$ is positive, while if the
lightcurve has a convex shape the average $\nudot$ will be
negative. This explains why $\nudot > 0$ in the non-flaring states and
$\nudot < 0$ in the flaring states. It suggests a direct influence of the
accretion rate on phase, which could be effectuated through the hot
spot position on the neutron star surface.  Extending this
interpretation to the average $\nudot$ over the entire outburst, we
also favor the interpretation of a moving hot spot for that long term
trend, discarding the hypothesis of a torque to explain the parabola
observed in the pulse phase residuals.

\citet{c08} also suggested that the lagging arrival times
observed during the flaring states cannot be explained with a torque
model, since they correspond to a sudden change from a spin up to a
spin down. These authors also suggested that motion of the hot spot
can be responsible for both the phase shifts and the increase of the
fractional amplitude during the flaring states. \citet{c08}
assumed a fixed position of the hot spot during the non-flaring
states.  However, it is unlikely that the hot spot is fixed on the
surface during the non-flaring state, as we have shown (see
\S~\ref{sec:nudot}) that the magnitude of the short-term $\nudot$ is
too large to be compatible with standard accretion theory.

\citet{Ibragimov09} recently proposed a receding disk as a
possible explanation for the timing noise and pulse profile
variability observed in the 2002 outburst of SAX J1808.4-3658.  In
this model the antipodal spot can be observed when the inner accretion
disk moves sufficiently far from the neutron star surface as a
consequence of decreasing flux.  We observed a strong overtone and
pulse phase drifts since the early stages of the outburst, when the
disk should be closest to the neutron star. Therefore it is not clear
whether our observations can be explained by this model or not, and
further investigations of the problem are required.

Two hot spots with different and variable intensities can produce
a phase shift and a changing pulsed amplitude, even if the
location of both hot spots is fixed on the neutron star surface
\citep{Burderi08}.
This possibility needs also further investigation since 
a self-consistent model has not yet been presented.

We observed (1) a relation between flux and time of arrivals for both
the flaring and non-flaring state (\S~\ref{flux-res}).  This relation
was consistent with being the same for the two states.  We also
observed (2) an anti-correlation between pulse fractional amplitudes
and time of arrivals during the flaring state.  Finally, (3) this
amplitude anti-correlation became stronger when using a long term
$\nudot$.  The amplitude anti-correlation was weak in the non-flaring
state, regardless of the $\nudot$.  In the context of a hot spot
motion model for the time of arrival variations, these findings
constrain the kinematics of this motion.\\

\citet{Lamb08} demonstrated that variations in the pulse
fractional amplitudes should be anti-correlated with their time of
arrival if the hot spot is close to the neutron star spin axis and the
hot spot wanders by a small amount in latitude.

\citet{Lamb08} showed that even a small displacement in longitude
of the emitting region, when close to the spin axis, produces a large
phase change, but no amplitude variation. A motion in latitude
produces both phase and amplitude changes due to the hot spot velocity
variation affecting Doppler boosting and aberration. An
anti-correlation between the pulse arrival times and the pulse
amplitudes would be an indicator of the above. Combining this with our
observational findings 1-3 above we conclude within the moving hot
spot model for the phase variation that:

1. the hot spot moves with flux in both flaring and non-flaring state,
since the relation between flux and arrival times is observed in both
cases and is consistent with being the same,

2. the amplitude anti-correlation in the flaring state implies an hot
spot moving in latitude.  The hot spot cannot move mainly in latitute
during the non-flaring state since a weak amplitude anti-correlation
is observed and the fractional amplitude changes by only a factor
$\approx 2$ in $106$ days.

3. The long term $\nudot$ must be related with a motion in longitude
since during the flaring state the amplitude anti-correlation becomes
much stronger when a $\nudot$ model is used to fit the time of arrivals.
This is also compatible with the non-flaring state, since the amplitude
anti-correlation remains weak with or without a $\nudot$.

4. Finally a motion in longitude during the flaring state or in
latitude during the non-flaring state is possible, but it has to be
small enough to preserve the observed flux and amplitude
anti-correlations.

The reason why the hot spot should drift mainly in longitude during
the non-flaring states and mainly in latidude during the flares might
be related with differences in the accretion flow process. A hot spot
motion has been observed in MHD simulations, with a complicated
dependence of the hot spot position on the misalignment angle between
magnetic field and rotation axis \citep{Romanova02, Romanova03,
Romanova04}.  As noted by \citet{Lamb08}, long term wandering of
the hot spot can be related with the structure in the inner
part of the accretion disk and therefore should track the long-term
changes of the accretion rate. The position of the hot spot on the
neutron star surface is expected to change rapidly and irregularly as
the accretion flow from the inner region of the accretion disk varies.
Further studies are required to couple our inferred hot spot
kinematics to physics and geometry of the accretion flow.  We note
that the fractional amplitudes can also change according to the hot
spot angular size and/or to the difference in temperature between the
hot spot and the neutron star surface.

The maximum observed sinusoidal fractional amplitude
(Fig.~\ref{fig:resid}:~$\approx 27\%$) can be explained if the hot
spot is slightly misaligned from the spin axis (colatitude $\simless
20^{\circ}$) with an inclination of the observer larger than $\approx
45^{\circ}$, or if the inclination of the observer is smaller than
$\approx 45^{\circ}$ but the spot has a large colatitude (see Figure 2
in \citealt{Lamb08}, note that we quote sinusoidal amplitudes while
they use rms amplitudes).

J1807 shows an anti-correlation between the second harmonic fractional
amplitude and the X-ray flux.  We observed a similar anti-correlation
in SAX J1808.4$-$3658 \citep{har08}.  This suggests the same
process as the origin of the anti-correlation in both pulsars.
\citet{har08} found that the anti-correlation was a signature
of the increasing asymmetry of the pulse profiles toward the end of
the outburst.  In J1807 the second harmonic is less often detected in
these late stages of the outburst. However, the lower count rates
late in the outburst lead to upper limits on the second harmonic there
that are sufficiently high that the explanation we proposed for J1808
\citep{har08} can still be valid for J1807 as well.

\section{Conclusions}

In this paper we analyzed the 2003 outburst of XTE J1807-294 and found
that the pulse frequency derivative previously reported in literature
is consistent with being part of a red noise process.  No significant
spin frequency derivative is detected when considering this red timing
noise as a source of uncertainty in the calculationn of statistical
uncertainties, and an upper limit of $5\times 10^{-14}\rm\,Hz\,s^{-1}$
can then be set for any spin frequency derivative. The average
accretion torque expected from standard accretion theory predicts a
long-term spin frequency derivative which is still compatible with the
derived upper limit and cannot therefore be excluded from current
observations.

We propose hot spot motion on the neutron star surface as a simpler
model able to explain all the observations reported in this work, as
well as the presence of a pulse frequency derivative.  If this
explanation is correct, similar flux and amplitude anti-correlations
should be observed in other AMXPs.

\acknowledgements{Acknowledgements: We would like to thank A.L. Watts,
P. Casella, D. Altamirano, F. Lamb, S. Boutloukos, C. Miller,
J. Poutanen and C. German\'{a} for stimulating discussions}

\end{document}